\documentclass[amstex,12pt]{article}
\usepackage{amsfonts}
\usepackage[utf8]{inputenc}
\usepackage{graphicx}
\usepackage[english]{babel}
\usepackage{amsthm}
\usepackage{xcolor}
\usepackage{float}
\usepackage{tikz}

\input tcilatex
\begin{document}

\author{Allard J., Universit\'{e} de Moncton, Canada, \and Champigny S.,
Universit\'{e} de Moncton, Canada, \and Choulakian V., Universit\'{e} de
Moncton, Canada, \and Mahdi S., University of the West Indies, Barbados,}
\title{TCA and TLRA: A comparison on  contingency tables and
compositional data}
\date{\today}
\maketitle

\begin{abstract}
There are two popular general approaches for the analysis and visualization
of a contingency table and a compositional data set: Correspondence analysis
(CA) and log ratio analysis (LRA). LRA includes two independently well
developed methods: association models and compositional data analysis. The
application of either CA or LRA to a contingency table or to compositional
data set includes a preprocessing centering step. In CA the centering step
is multiplicative, while in LRA  it is log bi-additive. A
preprocessed matrix is double-centered, so it is a residuel matrix; which
implies that it affects the final results of the analysis. This paper
introduces a novel index named the intrinsic measure of the quality of the
signs of the residuals (QSR) for the choice of the preprocessing, and
consequently of the method. The criterion is based on taxicab singular value
decomposition (TSVD) on which the package TaxicabCA in R is developed. We
present a minimal R script that can be executed to obtain the numerical
results and the maps in this paper. Three relatively small sized data sets
available freely on the web are used as examples.


\end{abstract}

\section{\textbf{Introduction}}

There are two popular general approaches for the analysis and visualization
of a contingency table or a compositional data set: Correspondence analysis
(CA) and log ratio analysis (LRA). LRA includes two independently well
developed methods: RC association models by Goodman (1991, 1996) and
compositional data analysis (CODA) by Aitchison (1986). Correspondence
analysis and log-ratio related methods are based on different invariance
principles: CA on Benz\'{e}cri's  \textit{distributional equivalence
principle}, RC association models on Yule's \textit{scale invariance
principle}, and CODA on Aitchison's \textit{subcompositional coherence
principle}. RC and CODA are mathematically speaking identical. Each of the
method, CA or LRA, includes a preprocessing-centering step of the data set.
In CA the preprocessing step is multiplicative, while in LRA it is log
bi-additive. A preprocessed contingency table or a compositional set is
double-centered, so it is a residual matrix which affects the subsequent
computations. Our aim is to introduce a simple intuitive criterion for the
choice of the preprocessing, and consequently of the method. The novel
criterion is the intrinsic measure of quality of the signs of the residuals
(QSR) by a principal dimension. For each principal dimension QSR is
calculated via taxicab singular value decomposition (TSVD) on which the
package TaxicabCA in R is developed. TaxicabCA in R is deleloped by Allard and Choulakian (2019). Three relatively small data sets, two
contingency tables and one compositional data set, available freely on the
web are used to explain the use of the package for the choice of the best
method between taxicab CA (TCA) or taxicab LRA (TLRA). The reference for correspondance analysis is  Benz\'{e}cri (1973). For a panoramic review of CA and its variants, see  Beh and Lombardo (2014).

This paper is organized as follows. In Section 2, we present an overview of
taxicab singular value decomposition (TSVD) and in  Section 3,  the
computation pertaining to  the methods TCA and TLRA.  Section 4 presents  the QSR index. In  Section 5, we
present three  examples and their analyses.  In Section 6, we present a minimal R script that
can be executed to obtain the numerical results and the maps in this paper.
Finally, we conclude in Section 7.

\section{An\ overview\ of\ taxicab\ singular value decomposition}

Consider a matrix $\mathbf{X}$\ of size $I\times J$ and $rank(\mathbf{X})=k$%
\textbf{.} Taxicab singular value decomposition (TSVD) of \textbf{X} is a
decomposition similar to SVD(\textbf{X}), see Choulakian (2006, 2016).

In TSVD the calculation of the dispersion measures $(\delta _{\alpha })$,
principal axes ($\mathbf{u}_{\alpha },\mathbf{v}_{\alpha })$ and principal
scores $(\mathbf{a}_{\alpha },\mathbf{b}_{\alpha })$ for $\alpha =1,...,k$
is done in an stepwise manner. We put $\mathbf{X}_{1}=\mathbf{X}=(x_{ij})$
and $\mathbf{X_{\alpha }}$ the residual matrix at the $\alpha $-th iteration
for $\alpha =1,...,k$.

The variational definitions of the TSVD at the $\alpha $-th iteration are

\begin{eqnarray}
\delta _{\alpha } &=&\max_{\mathbf{u\in
\mathbb{R}
}^{J}}\frac{\left\vert \left\vert \mathbf{X_{\alpha }u}\right\vert
\right\vert _{1}}{\left\vert \left\vert \mathbf{u}\right\vert \right\vert
_{\infty }}=\max_{\mathbf{v\in
\mathbb{R}
}^{I}}\ \frac{\left\vert \left\vert \mathbf{X_{\alpha }^{\prime }v}%
\right\vert \right\vert _{1}}{\left\vert \left\vert \mathbf{v}\right\vert
\right\vert _{\infty }}=\max_{\mathbf{u\in
\mathbb{R}
}^{J},\mathbf{v\in
\mathbb{R}
}^{I}}\frac{\mathbf{v}^{\prime }X\mathbf{_{\alpha }u}}{\left\vert \left\vert
\mathbf{u}\right\vert \right\vert _{\infty }\left\vert \left\vert \mathbf{v}%
\right\vert \right\vert _{\infty }},   \nonumber \\
&=&\max ||\mathbf{X_{\alpha }u||}_{1}\ \ \textrm{subject to }\mathbf{u}\in
\left\{ -1,+1\right\} ^{J},  \nonumber \\
&=&\max ||\mathbf{X_{\alpha }^{\prime }v||}_{1}\ \ \textrm{subject to }\mathbf{%
v}\in \left\{ -1,+1\right\} ^{I},  \nonumber \\
&=&\max \mathbf{v}^{\prime }\mathbf{X_{\alpha }u}\textrm{ \ subject to \ }%
\mathbf{u}\in \left\{ -1,+1\right\} ^{J},\mathbf{v}\in \left\{ -1,+1\right\}
^{I}.
\end{eqnarray}%
The $\alpha $-th principal axes are%
\begin{equation}
\mathbf{u}_{\alpha }\ =\arg \max_{\mathbf{u}\in \left\{ -1,+1\right\}
^{J}}\left\vert \left\vert \mathbf{X_{\alpha }u}\right\vert \right\vert _{1}%
\textrm{ \ \ and \ \ }\mathbf{v}_{\alpha }\ =\arg \max_{\mathbf{v}\in \left\{
-1,+1\right\} ^{I}}\left\vert \left\vert \mathbf{X_{\alpha }^{\prime }v}\right\vert
\right\vert _{1}\textrm{,}  
\end{equation}%
and the $\alpha $-th principal projections of the rows and the columns are
\begin{equation}
\mathbf{a}_{\alpha }=\mathbf{X_{\alpha }u}_{\alpha }\textrm{ \ and \ }\mathbf{b%
}_{\alpha }=\mathbf{X_{\alpha }^{\prime }v}_{\alpha }.  
\end{equation}%
Furthermore, the following relations are also useful%
\begin{equation}
\mathbf{u}_{\alpha }=sign(\mathbf{b}_{\alpha })\textrm{ \ and \ }\mathbf{v}%
_{\alpha }=sign(\mathbf{a}_{\alpha }),  
\end{equation}%
where $sign(.)$ is the coordinatewise sign function, $sign(x)=1$ \ if \ $%
x>0, $ \ and \ $sign(x)=-1$ \ if \ $x\leq 0.$

The $\alpha $-th taxicab dispersion measure $\delta _{\alpha }$ can be
represented in many different ways%
\begin{eqnarray}
\delta _{\alpha }\ &=&\left\vert \left\vert \mathbf{X_{\alpha }u}_{\alpha
}\right\vert \right\vert _{1}=\left\vert \left\vert \mathbf{a}_{\alpha
}\right\vert \right\vert _{1}=\mathbf{a}_{\alpha }^{\prime }\mathbf{v}%
_{\alpha },   \\
&=&\left\vert \left\vert \mathbf{X_{\alpha }^{\prime }v}_{\alpha
}\right\vert \right\vert _{1}=\left\vert \left\vert \mathbf{b}_{\alpha
}\right\vert \right\vert _{1}=\mathbf{b}_{\alpha }^{\prime }\mathbf{u}%
_{\alpha }.  \nonumber
\end{eqnarray}%
The $(\alpha +1)$-th residual matrix is
\begin{equation}
\mathbf{X_{\alpha +1}}=\mathbf{X_{\alpha }-a}_{\alpha }\mathbf{b}_{\alpha
}^{\prime }/\delta _{\alpha }.  
\end{equation}%
An interpretation of the term $\mathbf{a}_{\alpha }\mathbf{b}_{\alpha
}^{\prime }/\delta _{\alpha }$ in (6) is that, it represents the best rank-1
approximation of the residual matrix $\mathbf{X_{\alpha }}$, in the sense of
the taxicab matrix norm (1).

Thus TSVD($\mathbf{X}$) corresponds to the bilinear decompostion

\QTP{}
\begin{equation}
x_{ij}=\sum_{\alpha =1}^{k}a_{\alpha }(i)b_{\alpha }(j)/\delta _{\alpha },
\end{equation}%
a decomposition similar to SVD, but where the vectors $(\mathbf{a}_{\alpha },%
\mathbf{b}_{\alpha })$ for $\alpha =1,...,k$ are conjugate, that is%
\begin{eqnarray}
\mathbf{a}_{\alpha }^{\prime }\mathbf{v}_{\beta } &=&\mathbf{a}_{\alpha
}^{\prime }sign(\mathbf{a}_{\beta })   \\
&=&\mathbf{b}_{\alpha }^{\prime }\mathbf{u}_{\beta }=\mathbf{b}_{\alpha
}^{\prime }sign(\mathbf{b}_{\beta })  \nonumber \\
&=&0\textrm{ for }\beta \geq \alpha +1.  \nonumber
\end{eqnarray}

\QTP{}
In the package TaxicabCA in R, the calculation of the principal component
weights, $\mathbf{u}_{\alpha }$ and $\mathbf{v}_{\alpha },$ are accomplished
by three algorithms. The first one, based on complete enumeration equation
(2), is named \textit{exhaustive}. The second one, based on iterating the
transition formulae (3,4), is named \textit{criss-cross}. The third one is
based on the genetic algorithm named \textit{genetic}.

\section{TCA and TLRA }

Let $\mathbf{N=(}n_{ij})$ be a contingency table or a compositional data set
of size $I\times J$, where $n_{ij}\geq 0$. Let $\mathbf{P=N/}t=(p_{ij})$ be
the associated correspondence matrix, where $t=\sum_{i=1}^{I}%
\sum_{j=1}^{J}n_{ij}.$ We define as usual $p_{i\ast }=\sum_{j=1}^{J}p_{ij}$
and $p_{\ast j}=\sum_{i=1}^{I}p_{ij}$ the row and column marginals,
respectively. We present the three steps of calculation necessary in TCA and
TLRA.

\subsection{TCA}

Step 1: Center the data:

\begin{eqnarray}
\mathbf{X}_{1} &=&(X_{1}(i,j))   \\
&=&(p_{ij}-p_{i\ast }p_{\ast j}).  \nonumber
\end{eqnarray}

Step 2: Calculate TSVD($\mathbf{X}_{1})$%
\begin{equation}
p_{ij}-p_{i\ast }p_{\ast j}=\sum_{\alpha =1}^{k}a_{\alpha }(i)b_{\alpha
}(j)/\delta _{\alpha }.  
\end{equation}

Step 3: Calculate TCA($\mathbf{P)}$ by dividing each term in (10) by $%
p_{i\ast }p_{\ast j}$%
\begin{equation}
\frac{p_{ij-}p_{i\ast }p_{\ast j}}{p_{i\ast }p_{\ast j}}=\sum_{\alpha
=1}^{k}f_{\alpha }(i)g_{\alpha }(j)/\delta _{\alpha },  
\end{equation}%
where evidently $f_{\alpha }(i)=$ $a_{\alpha }(i)/p_{i\ast }$ and $g_{\alpha
}(j)=b_{\alpha }(j)/p_{\ast j}.$ We name $(a_{\alpha }(i),b_{\alpha }(j))$
TCA contribution scores. Similarly, $(f_{\alpha }(i),g_{\alpha }(j))$ are
named TCA principal scores.

\subsection{TLRA}

We obtain TLRA, by weighing each row and column uniformly. Then we procceed according to the following steps:
\bigskip

Step 1: Center the log data $G_{ij}=\log (p_{ij})$:
\begin{equation}
X_{1}(i,j)=G_{ij}-G_{i\ast }-G_{\ast j}+G_{\ast \ast },  
\end{equation}%
where $G_{i\ast }=\sum_{j=1}^{J}G_{ij}/J,$ $G_{\ast j}=\sum_{i=1}^{I}G_{ij}/I
$ and $G_{\ast \ast }=\sum_{j=1}^{J}\sum_{i=1}^{I}G_{ij}/(IJ).$ This is
equation 2.2.1 in Goodman(1991) or equation 5 in Goodman(1996).

Step 2: Calculate TSVD$(\mathbf{X}_{1})$%
\begin{equation}
X_{1}(i,j)=\sum_{\alpha =1}^{k}a_{\alpha }(i)b_{\alpha }(j)/\delta _{\alpha
}.  
\end{equation}

Step 3: Calculate TLRA$(\mathbf{P)}$ by dividing each term in (13) by $1/(IJ)
$%
\begin{equation}
\frac{G_{ij}-G_{i\ast }-G_{\ast j}+G_{\ast \ast }}{1/(IJ)}=\sum_{\alpha
=1}^{k}f_{\alpha }(i)g_{\alpha }(j)/\delta _{\alpha },  
\end{equation}%
where evidently $f_{\alpha }(i)=$ $I\ a_{\alpha }(i)$ and $g_{\alpha }(j)=J\
b_{\alpha }(j).$We name $(a_{\alpha }(i),b_{\alpha }(j))$ TLRA contribution
scores. Similarly, $(f_{\alpha }(i),g_{\alpha }(j))$ are named TLRA
principal scores.

\subsection{Facts}

Fact 1: In both  methods the matrix $\mathbf{X}_{1}=(X_{1}(i,j))$ is
double-centered%
\[
\sum_{i=1}^{I}X_{1}(i,j)=\sum_{j=1}^{J}X_{1}(i,j)=0.
\]

Fact 2: The set of scores ($a_{\alpha }(i))$ and ($b_{\alpha }(j))$, besides
satisfying (5) and (8) are centered%
\[
\sum_{i=1}^{I}a_{\alpha }(i)=\sum_{j=1}^{J}b_{\alpha }(j)=0\textrm{ \ \ \ for }%
\alpha =1,...,k.
\]

Fact 3: Let $I_{1}=\left\{ 1,...,I\right\} $ and $J_{1}=\left\{
1,...,J\right\} ;$ and $S\cup \overline{S}=I_{1}$ be the partition of $I_{1}$%
, and $T\cup \overline{T}=J_{1}$ be the partition of $J_{1},$ such that $%
S=\left\{ i:a_{\alpha }(i)>0\right\} $ and $T=\left\{ j:b_{\alpha
}(j)>0\right\} .$  $ \overline{S}$ and  $\overline{T}$ are  the complements of $S$  and $T$, respectively. Let also,
\begin{eqnarray*}
X_{m+1}(i,j) &=&X_{1}(i,j)-\sum_{\alpha =1}^{m}a_{\alpha }(i)b_{\alpha
}(j)/\delta _{\alpha }\textrm{\ \ \ for\ \ }m=1,...,k-1, \\
&=&X_{m}(i,j)-a_{m}(i)b_{m}(j)/\delta _{m}
\end{eqnarray*}%
be $(m+1)$th residual matrix. Besides (5), the taxicab dispersion $\delta
_{m}$ will additionally be related to the contribution scores $a_{m}(i)$ and
$b_{m}(j)$ in (10,14) by the following useful equations, see Choulakian and
Abou-Samra (2020):

\begin{eqnarray}
\delta _{m}/2 &=&\sum_{i\in S}a_{m}(i)=-\sum_{i\in \overline{S}}a_{m}(i)
 \\
&=&\sum_{j\in T}b_{m}(j)=-\sum_{j\in \overline{T}}b_{m}(j);  \nonumber
\end{eqnarray}%
which tells that the principal dimensions are \textit{balanced}. Furthermore%
\begin{eqnarray}
\delta _{m}/4 &=&\sum_{(i,j)\in S\times T}X_{m}(i,j)=\sum_{(i,j)\in
\overline{S}\times \overline{T}}X_{m}(i,j)   \\
&=&-\sum_{(i,j)\in \overline{S}\times T}X_{m}(i,j)=-\sum_{(i,j)\in S\times
\overline{T}}X_{m}(i,j);  \nonumber
\end{eqnarray}%
which tells that the $m$th principal dimension divides the residual data
matrix $X_{m}$ into 4 \textit{balanced} quadrants.

In both methods, the symmetric maps are obtained by plotting $(f_{\alpha
}(i),\ f_{\beta }(i))$ or $(g_{\alpha }(j),\ g_{\beta }(j))$ for $\alpha
\neq \beta .$

\section{Quantifying the intrinsic quality of a taxicab principal axis}

We briefly review the quality of measures of a principal dimension in the
Euclidean framework, then within the Taxicab framework.

\subsection{Euclidean framework}

Within the Euclidean framework a common used measure of the quality of a
principal dimension $\alpha $ of the residual matrix $\mathbf{X}_{1}$
described in (13), is the proportion of variance explained (or inertia in
the case of CA)%
\begin{eqnarray*}
\tau _{1}(\alpha ) &=&\%({\textrm{explained total variance by dimension }}\alpha
) \\
&=&100\frac{\sigma _{\alpha }^{2}}{\sum_{\beta =1}^{k}\sigma _{\beta }^{2}}\
\ \ \ \textrm{for\ \ }\alpha =1,...,k \\
&=&100\frac{\sigma _{\alpha }^{2}}{\sum_{(i,j)}|X_{1}(i,j)|^{2}}.
\end{eqnarray*}%
Another variant is
\begin{eqnarray*}
\tau _{2}(\alpha ) &=&\%(\textrm{explained residual variance by dimension }%
\alpha ) \\
&=&100\frac{\sigma _{\alpha }^{2}}{\sum_{\beta =\alpha }^{k}\sigma _{\beta
}^{2}}\ \ \ \ \textrm{for\ \ }\alpha =\beta ,...,k \\
&=&100\frac{\sigma _{\alpha }^{2}}{\sum_{(i,j)}|X_{\alpha }(i,j)|^{2}}.
\end{eqnarray*}%
Note that $\tau _{1}(\alpha )$ and $\tau _{2}(\alpha )$ are extrinsic
measures of quality of the residuals in the residual matrix $\mathbf{X}%
_{\alpha }$, because they compare the intrinsic dispersion of a principal
axis $\sigma _{\alpha }^{2}$ to the total dispersion $\sum_{\alpha
=1}^{k}\sigma _{\alpha }^{2}$ or to the partial residual dispersion $%
\sum_{\beta =\alpha }^{k}\sigma _{\beta }^{2}$. Furthermore, when $\tau _{1}(\alpha )$ and $\tau _{2}(\alpha )$  are expressed in proportion, we have the following evident result that should be compared with Lemma 2.\bigskip

\textbf{Lemma 1}:
\begin{itemize}
\item{a)}  $1>\tau _{i}(\alpha )$ for $i=1,2$ and  $\alpha
=1,...,k-1$.
\item{b)} For $\alpha =k,1=\tau _{2}(\alpha ).$
\end{itemize}
\subsection{Taxicab framework}

The Taxicab variant of $\tau _{2}$ is particularly adapted in TSVD%
\[
QSR_{\alpha }=\frac{\delta _{\alpha }}{\sum_{(i,j)}|X_{\alpha }(i,j)|},
\]%
which we will interpret as a new intrinsic measure of quality of the signs
of the residuals in the residual matrix $\mathbf{X}_{\alpha }$ for $\alpha
=1,...,k$. As usual $|a|$ designates absolute value of the real number $a.$

Let $S\cup \overline{S}=I_{1}$ be the optimal principal axis partition of $%
I_{1}$, and similarly $T\cup \overline{T}=J_{1}$ be the optimal principal
axis partition of $J_{1},$ such that $S=\left\{ i:a_{\alpha }(i)>0\right\} =$
$\left\{ i:v_{\alpha }(i)>0\right\} $and $T=\left\{ j:b_{\alpha
}(j)>0\right\} =\left\{ j:u_{\alpha }(j)>0\right\} $ by (4). Thus the data
set is divided into four quadrants. Based on the equations (16), we define a
new index quantifying the quality of the signs of the residuals in each
quadrant of the $\alpha $th residual matrix $\mathbf{X}_{\alpha }$ for $%
\alpha =1,...,k$.\bigskip

\textbf{Definition}:\ For $\alpha =1,...,k-1,$ an intrinsic measure of the
quality of the signs of the residuals in the quadrant $E\times F\subseteq
I_{1}\times J_{1}$ is
\begin{eqnarray*}
QSR_{\alpha }(E,F) &=&\frac{\sum_{(i,j)\in E\times F}X_{\alpha }(i,j)}{%
\sum_{(i,j)\in E\times F}|X_{\alpha }(i,j)|} \\
&=&\frac{\delta _{\alpha }/4}{\sum_{(i,j)\in E\times F}|X_{\alpha }(i,j)|}%
\end{eqnarray*}%
for $E=S$ and $\overline{S},$ and, $F=T$ and $\overline{T}.$\ The second  right-hand side in the above equation derives from equation $ (16)$. \bigskip

We have the following easily proved\bigskip

\textbf{Lemma 2}: a) For $\alpha =1,...,k-1,$ $QSR_{\alpha }=1$ if and only
if $QSR_{\alpha }(S,T)=QSR_{\alpha }(\overline{S},\overline{T})=-QSR_{\alpha
}(S,\overline{T})=-QSR_{\alpha }(\overline{S},T)=1.$

b) For $\alpha =k,$ $QSR_{\alpha }=1.\bigskip $

The interpretation of $QSR_{\alpha }(E,F)=\pm 1$ is that in the quadrant $%
E\times F$ the residuals have one sign; and this is a signal for very
influential cells or columns or rows. Example 1 explains this fact. So Lemma
2 provides a necessary and sufficient condition for $QSR_{\alpha }=1,$ which
is not true for $\tau _{1}(\alpha )$ and $\tau _{2}(\alpha )$. Geometry
plays its unique role.\bigskip

\textbf{Notation}: $QSR_{\alpha }(+)=\left\{ QSR_{\alpha }(S,T),QSR_{\alpha
}(\overline{S},\overline{T})\right\} $ and $QSR_{\alpha }(-)=\left\{
QSR_{\alpha }(S,\overline{T}),QSR_{\alpha }(\overline{S},T)\right\}
.\bigskip $

\textbf{Remark: }The computation of the elements of $QSR_{\alpha }(+)$ and $%
QSR_{\alpha }(-)$ are done easily in the following way. We note that the $%
\alpha $th principal axis can be written as%
\[
\mathbf{u}_{\alpha }=\mathbf{u}_{\alpha +}+\mathbf{u}_{\alpha -},
\]%
where $\mathbf{u}_{\alpha +}=(\mathbf{u}_{\alpha }+\mathbf{1}_{J})/2$ and $%
\mathbf{u}_{\alpha -}=(\mathbf{u}_{\alpha }-\mathbf{1}_{J})/2;$ similarly
\[
\mathbf{v}_{\alpha }=\mathbf{v}_{\alpha +}+\mathbf{v}_{\alpha -},
\]%
where $\mathbf{v}_{\alpha +}=(\mathbf{v}_{\alpha }+\mathbf{1}_{I})/2$ and $%
\mathbf{v}_{\alpha -}=(\mathbf{v}_{\alpha }-\mathbf{1}_{I})/2$, where $%
\mathbf{1}_{I}$ designates a column vector of 1's of size $I.$ So%
\[
QSR_{\alpha }(S,T)=\frac{\delta _{\alpha }/4}{\mathbf{v}_{\alpha +}^{\prime
}abs(\mathbf{X}_{\alpha })\mathbf{u}_{\alpha +}},
\]%
\[
QSR_{\alpha }(\overline{S},\overline{T})=\frac{\delta _{\alpha }/4}{\mathbf{v%
}_{\alpha -}^{\prime }abs(\mathbf{X}_{\alpha })\mathbf{u}_{\alpha -}},
\]%
\[
QSR_{\alpha }(S,\overline{T})=\frac{\delta _{\alpha }/4}{\mathbf{v}_{\alpha
-}^{\prime }abs(\mathbf{X}_{\alpha })\mathbf{u}_{\alpha +}},
\]%
\[
QSR_{\alpha }(\overline{S},T)=\frac{\delta _{\alpha }/4}{\mathbf{v}_{\alpha
+}^{\prime }abs(\mathbf{X}_{\alpha })\mathbf{u}_{\alpha -}},
\]%
where $abs(\mathbf{X}_{\alpha })=(|X_{\alpha }(i,j)|).$

\section{Examples}

Here we present the analysis of two contingency tables and one compositional
data set.

\subsection{Xlstat demoCA count data set}

Table 1 is a small data set of size $7\times 4,$ as the title suggests to
introduce CA in the software Xlstat available by a google search on the web.

\begin{tabular}{l|llll}
\multicolumn{5}{l}{\textbf{Table 1: Xlstat demoCA count table.}} \\ \hline
\multicolumn{5}{c}{Attribute} \\
Age & Bad & Average & Good & VeryGood \\ \hline
16-24 & $69$ & $49$ & $48$ & $41$ \\
25-34 & $148$ & $45$ & $14$ & $22$ \\
35-44 & $170$ & $65$ & $12$ & $29$ \\
45-54 & $159$ & $57$ & $12$ & $28$ \\
55-64 & $122$ & $26$ & $6$ & $18$ \\
65-74 & $106$ & $21$ & $5$ & $23$ \\
75+ & $40$ & $7$ & $1$ & $14$ \\ \hline
\end{tabular}
\newline
\vspace*{0.5cm}
\newline
\begin{tabular}{l|l|lll||l|}
\multicolumn{6}{l}{\textbf{Table 2: 10}$^{3}\times $\textbf{Xlstat demoCA
count table TCA centered.}} \\ \hline
\multicolumn{5}{c||}{Attribute} & row \\
Age & Bad & Average & Good & VeryGood & sum \\ \hline
16-24(a) & $-40.66$ & $5.76$ & $24.36$ & $10.54$ & $0$ \\ \cline{2-6}
25-34(b) & $7.84$ & $-0.42$ & $-1.87$ & $-5.55$ & $0$ \\
35-44(c) & $3.27$ & $7.43$ & $-5.85$ & $-4.86$ & $0$ \\
45-54(d) & $4.00$ & $4.47$ & $-4.78$ & $-3.69$ & $0$ \\
55-64(e) & $13.87$ & $-6.06$ & $-4.73$ & $-3.08$ & $0$ \\
65-74(f) & $9.60$ & $-7.25$ & $-4.56$ & $2.22$ & $0$ \\
75+(g) & $2.07$ & $-3.93$ & $-2.56$ & $4.42$ & $0$ \\ \hline\hline
column\ sum & $0$ & $0$ & $0$ & $0$ & $0$ \\ \hline
\end{tabular}
\newline
\vspace*{0.5cm}
\newline
\begin{tabular}{l|ll|ll||l|}
\multicolumn{6}{l}{\textbf{Table 3: 10}$^{3}\times $\textbf{Xlstat demoCA
count table TLRA centered.}} \\ \hline
\multicolumn{5}{c||}{Attribute} & row \\
Age & Bad & VeryGood & Good & Average & sum \\ \hline
16-24 & $-0.9994$ & $-0.0309$ & $1.1679$ & $-0.1377$ & $0$ \\
25-34 & $0.0579$ & $-0.3592$ & $0.2299$ & $0.0714$ & $0$ \\
35-44 & $0.0394$ & $-0.2401$ & $-0.0813$ & $0.2820$ & $0$ \\
45-54 & $0.0308$ & $-0.2168$ & $-0.0230$ & $0.2090$ & $0$ \\ \hline
55-64 & $0.3122$ & $-0.1124$ & $-0.1699$ & $-0.0298$ & $0$ \\
65-74 & $0.2444$ & $0.2055$ & $-0.2794$ & $-0.1705$ & $0$ \\
75+ & $0.3146$ & $0.7538$ & $-0.8441$ & $-0.2244$ & $0$ \\ \hline\hline
col\ sum & $0$ & $0$ & $0$ & $0$ & $0$ \\ \hline
\end{tabular}
\newline
\vspace*{0.5cm}
\newline

Tables 2 and 3 display two different ways of double centering the data in
TCA and LRA. It is evident that corresponding residuals in  both entries
can be different, thus producing probably different quality maps, even
though we shall use the same algorithm in the sequel.

Let us describe in a nutshell what TSVD does on the TCA residuals in Table
2. Our aim is to partition the TCA residuals into four quadrants by
permuting the rows and the columns in such a way that the signs in each
quadrant are mostly constant; or equivalently by maximizing $QSR_{1}$ index.
The number of nontrivial partitions is: $(2^{I}-2)(2^{J}-2).$ For TCA , the
partition that maximizes $QSR_{1}=81.43\%$ is delineated in Table 2, where: $%
S=\left\{ VeryGood,\ Good,\ Average\right\} $ and $\overline{S}=\left\{
Bad\right\} ,$ and, $T=\left\{ a\right\} $ and $\overline{T}=\left\{
b,c,d,e,f,g\right\} .$ Note that
\[
QSR_{1 }(+)=\left\{ QSR_{1 }(S,T)=100,QSR_{1 }(\overline{S},%
\overline{T})=100\right\}
\]%
as reported in Table 4: because the residuals in the quadrant $S\times T$
are all positive $\left\{ 5.76,\ 24.36,\ 10.54\right\} ,$ and, the residuals
in the quadrant $\overline{S}\times \overline{T}$ are all positive $\left\{
7.84,\ 3.27,\ 4,\ 13.87,\ 9.6,\ 2.07\right\} .\ $Similarly as reported in
Table 4,
\[
QSR_{1}(-)=\left\{ QSR_{1}(S,\overline{T})=-52.29,QSR_{1}(\overline{S}%
,T)=-100\right\} .
\]%
$QSR_{1}(S,\overline{T})=-52.29$ is quite low, because the quadrant $S\times
\overline{T}$ has 18 residuals of which four have positive sign and 14
negative sign; while $QSR_{1}(\overline{S},T)=-100$, because the quadrant $%
\overline{S}\times T$ is a singleton $\left\{ -40.66\right\} $; so the singleton cell produces one heavyweight column, ``bad" and one heavyweight row, ``16-24", but it is not a heavyweight cell, because its weight does not go to infinity as discussed in Choulakian (2008).
Furthermore, we also note that the first taxicab dispersion $\delta _{1}=4\
\frac{|-40.66|}{1000}=0.1626,$ because of (16). The column $Bad$ and the
age group $16-24$ dominate the first dimension of the TCA map  in  Figure \ref{smith:fig1b}  .

Note that the optimal partitions in Table 2 and Table 3 are different.

Figure \ref{smith:fig1b} displays both  TCA and TLRA maps of the data with distinct colors for age category and modality. We note the
following two facts. First, the TCA and TLRA maps are quite different;
second, the TLRA map in  Figure \ref{smith:fig1b} is much more interpretable than the
corresponding TCA map, because the age groups are ordered on the first axis
in the TLRA map.

Table 4 displays the intrinsic measures of quality of the signs of the
residuals, $QSR$ values, for the first two principal dimensions for TCA and
TLRA. TLRA values are $QSR_{1}=87.69$ and $QSR_{2}=94.90,$ which are higher
than the corresponding TCA values $QSR_{1}=81.43$ and $QSR_{2}=86.79$. So
the $QSR$ values confirm what we saw visually.

\begin{tabular}{lllll}
\multicolumn{5}{l}{\textbf{Table 4: QSR (\%) of Xlstat demoCA data.}} \\
\hline
&  & TCA &  &  \\
$\alpha $ & $QSR_{\alpha }(+)$ & $QSR_{\alpha }(-)$ & $QSR_{\alpha }$ & $%
\delta _{\alpha }$ \\ \hline
1 & (\textbf{100,\ 100}) & (\textbf{-100, }-52.29,\ \textbf{)} & \textbf{%
81.43} & 0.1626 \\
2 & (\textbf{100},\ 83.74\textbf{)} & (\textbf{-100,\ }-70.69) & \textbf{%
86.79} & 0.0545 \\ \hline\hline
&  & TLRA &  &  \\ \hline
$\alpha $ & $QSR_{\alpha }(+)$ & $QSR_{\alpha }(-)$ & $QSR_{\alpha }$ & $%
\delta _{\alpha }$ \\ \hline
1 & (78.02\textbf{,\ 88.43}) & (\textbf{-100},\ -87.02\textbf{)} & \textbf{%
87.69} & 6.8725 \\ \hline
2 & (90.76,\ \textbf{99.44)} & (\textbf{-99.44,\ }-90.76) & \textbf{94.90} &
4.390 \\ \hline
\end{tabular}

The last column in Table 4 displays the first two taxicab dispersion values $%
\delta _{\alpha }$ for $\alpha =1,2$ for the two methods, which are not
comparable.

\begin{table}
\begin{tabular}{c}
  \includegraphics[width=19.3cm]{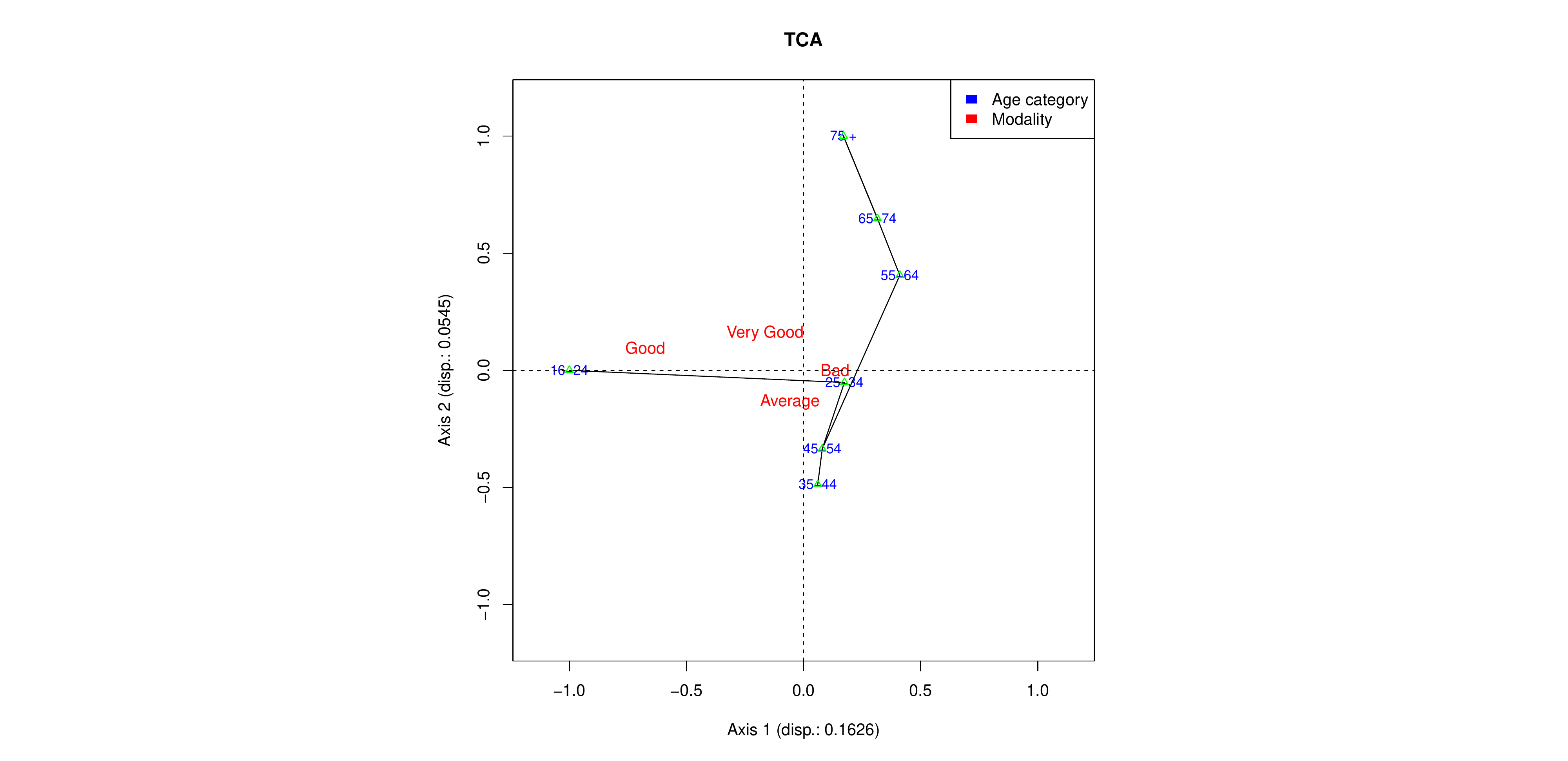}\\
  \includegraphics[width=19.3cm]{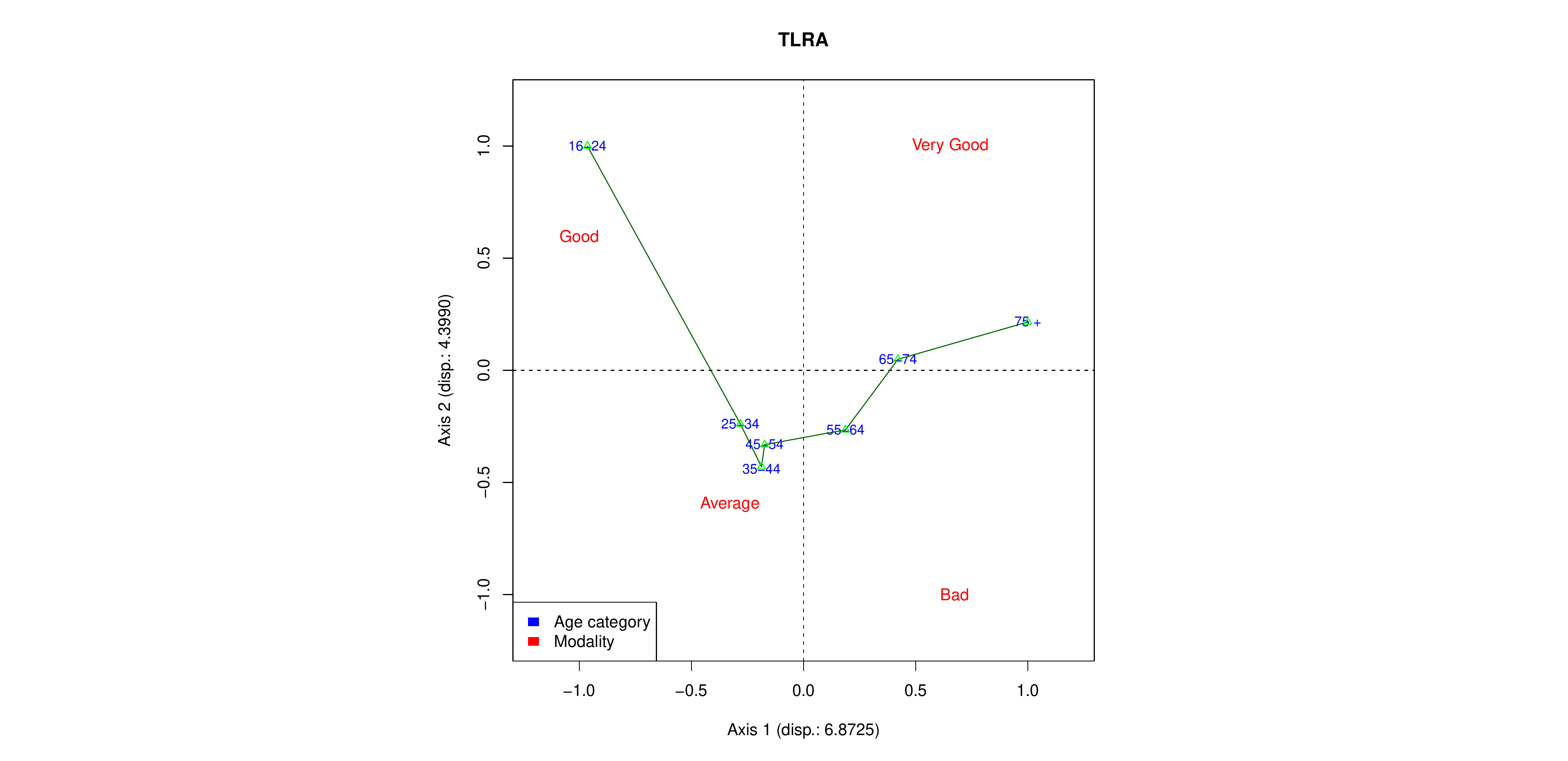} 
\end{tabular}
\caption{\label{smith:fig1b}  \textbf{ {\it Maps of DemoCA data.}}} 
\end{table}

\subsection{English authors count data}

This is a sparse contingency table cross-classifying known english authors
according to their geographical origins described by 50 counties and 12
periods of length 25 years extending from 1300-1600; it can be found in
Genet (2002). For TLRA computations we have added 1 to all counts, a
procedure suggested by Tukey (1977, p. 257), which keeps the re-expressed
log-counts nonnegative. Figure $\ref{smith:fig2}$ displays  the TCA and TLRA maps  of the data. It is showing
only the 12 periods. We note the following facts: The TCA map is more
interpretable than the TLRA map. We interpret the TCA map in the following
way: by grouping the first seven periods extending from $1300$ to $1475$, we obtain
a parabolic structure on the principal plane which shows the evolution of
the geographical origins of the authors described by the counties. For
further details concerning TCA of sparse contingency tables, refer to
Choulakian (2017).

Table 5 presents the QSR values for the first four principal dimensions: We
choose the TCA method because its $QSR_{1}=54.81$ and $QSR_{2}=50.29$
indices are higher by 10\% than the corresponding TLRA values $QSR_{1}=43.28$
and $QSR_{2}=40.97$.

\begin{tabular}{lllll}
\multicolumn{5}{l}{\textbf{Table 5: QSR (\%) of Authors data for the first 4
dimensions.}} \\ \hline\hline
&  & TCA &  &  \\
$\alpha $ & $QSR_{\alpha }(+)$ & $QSR_{\alpha }(-)$ & $QSR_{\alpha }$ & $%
\delta _{\alpha }$ \\ \hline
1 & (\textbf{73.56,\ }40.06) & (\textbf{-84.19},\ -44.35\textbf{)} & \textbf{%
54.81} & 0.2380 \\
2 & (\textbf{67.84},\ 36.64\textbf{)} & (\textbf{-67.31,\ }-44.15) & \textbf{%
50.29} & 0.1975 \\
3 & (43.18,\ 60.27) & (-49.54, -57.58) & 51.74 & 0.1753 \\
4 & (42.21, 50.78) & (-44.84, -59.85) & 48.5 & 0.1373 \\ \hline\hline
&  & TLRA &  &  \\
$\alpha $ & $QSR_{\alpha }(+)$ & $QSR_{\alpha }(-)$ & $QSR_{\alpha }$ & $%
\delta _{\alpha }$ \\ \hline
1 & (\textbf{50.56,\ }38.14) & (\textbf{-48.87},\ -38.53\textbf{)} & \textbf{%
43.28} & 93.6699 \\
2 & (40.99,\ \textbf{45.24)} & (-33.02\textbf{,\ -47.97)} & \textbf{40.97} &
78.2741 \\
3 & (31.30,\ 59.29) & (-45.79, -46.63) & 43.43 &  74.8385 \\
4 & (55.55, 33.78) & (-38.67, -51.99) & 43.15 &  66.8444 \\ \hline
\end{tabular}

\begin{table}
\begin{tabular}{c}
  \includegraphics[width=19.3cm]{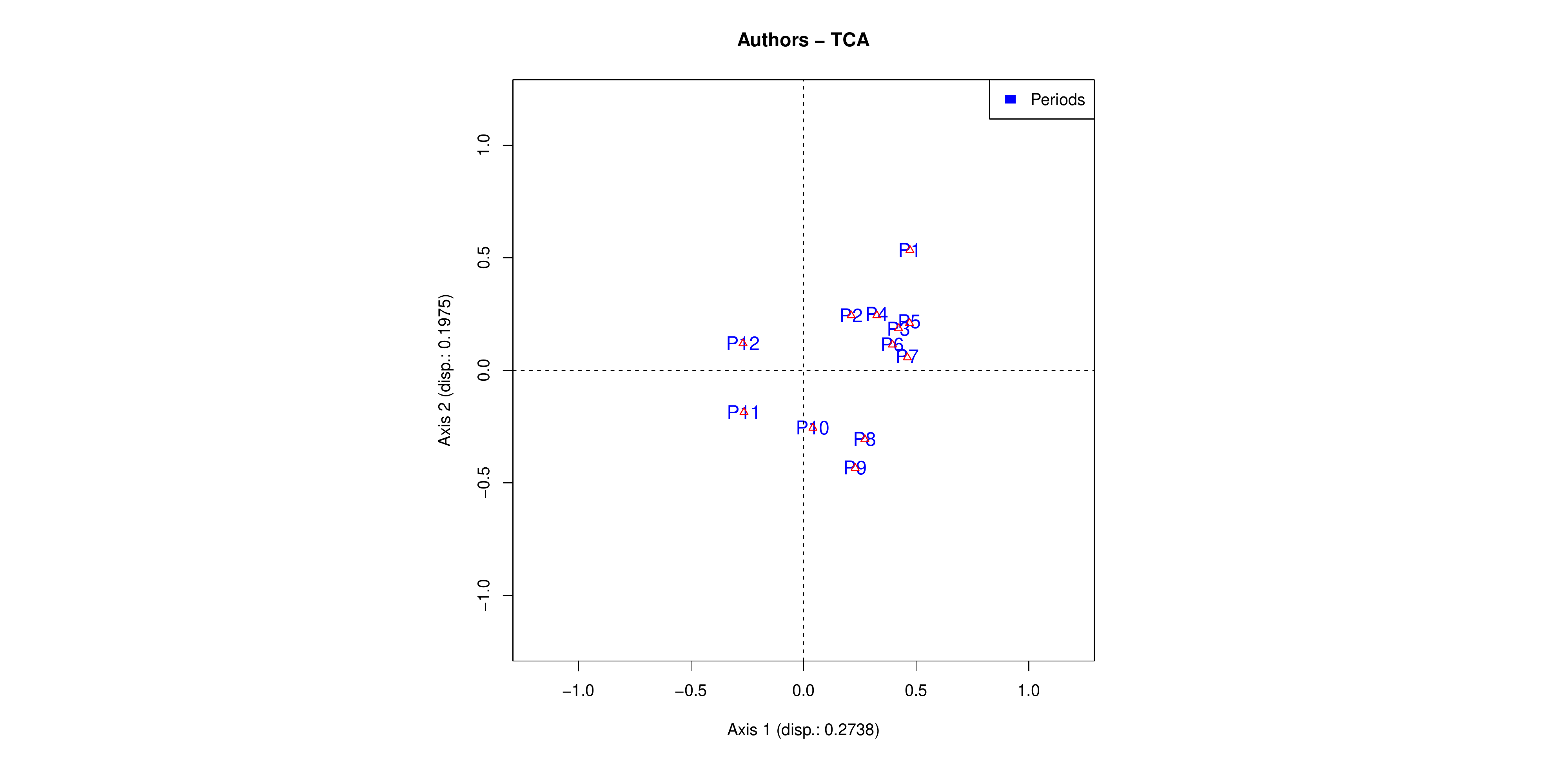}\\
 \includegraphics[width=19.3cm]{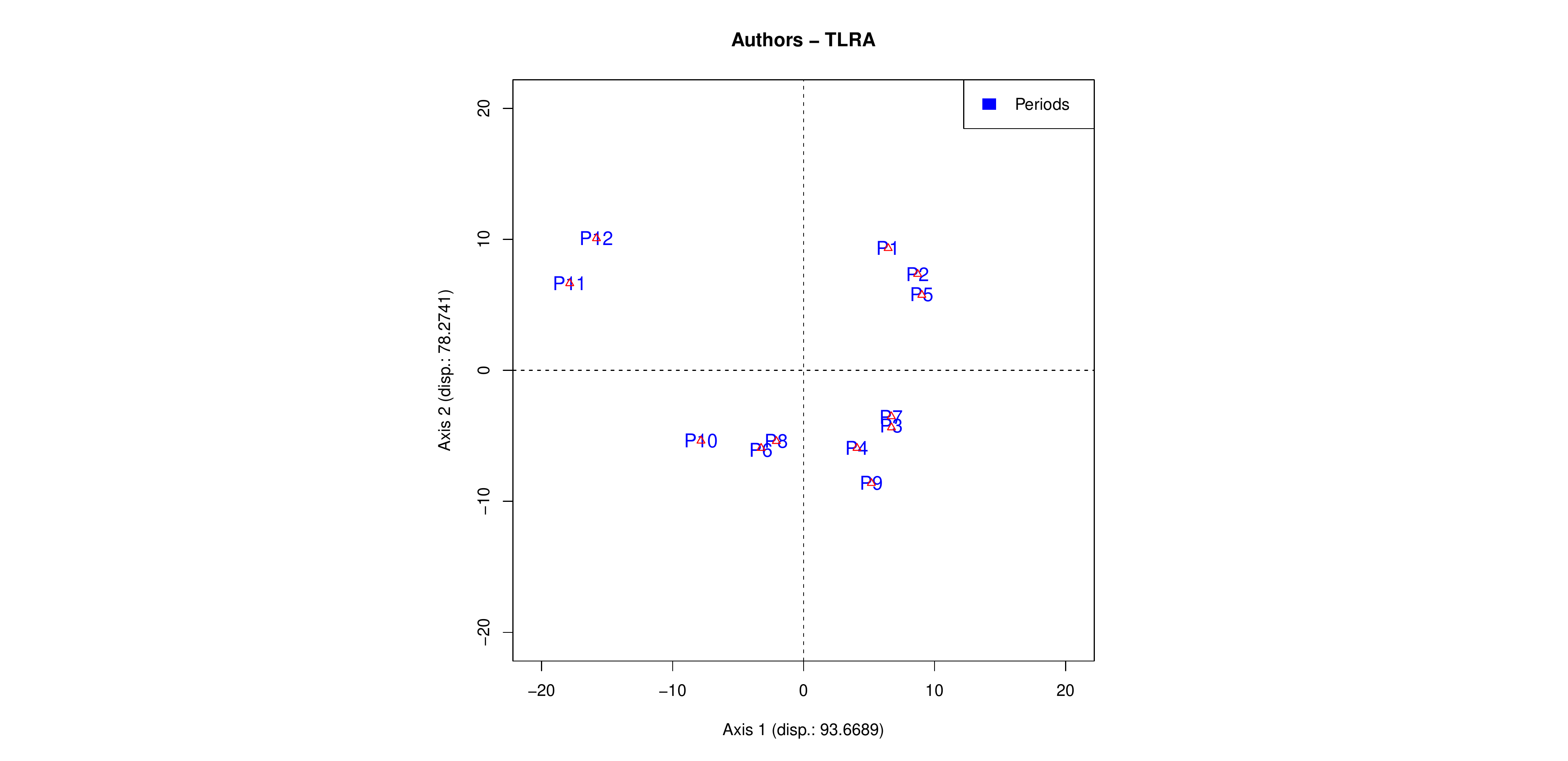}
\end{tabular}
\caption{\label{smith:fig2}  \textbf{ {\it Maps of Authors  data.}}}
\end{table}

\subsection{Food compositional data}

The food compositional data set is of size 25 by 9 and analyzed quite in
detail by CODA-LRA in Pawlowsky-Glahn and Egozcue (2011). These data are
percentages of consumption of 9 different kinds of food in 25 countries in
Europe in the early eighties. The 9 different kinds of food are: red meat
(RM); white meat (WM); fish (F); eggs (E); milk (M); cereals (C); starch
(S); nuts (N); fruit and vegetables (FV). The 25 countries are divided into
16 western (w) and 9 eastern (e) countries. It is evident that in Table 6,
TCA $QSR_{1}=77.89$ value is significantly higher than the TLRA $%
QSR_{1}=68.69$, so we choose TCA over TLRA. In Figure \ref{smith:fig3}  are displayed the TCA and TLRA maps where
 the 9 food kinds  are represented by their symbols and the 25 countries by their
symbols eastern (e) or western (w). The TCA map discriminates much better
the eastern and the western countries than the TLRA map: All eastern
countries (except 1 located in the first quadrant) are clustered in the
third quadrant. We also note that TLRA map is very similar to LRA map in
Pawlowsky-Glahn and Egozcue (2011).
\newline
\vspace*{0.1 cm}
\newline

\begin{tabular}{lllll}
\multicolumn{5}{l}{\textbf{Table 6: QSR (\%) of Food data for the first 4
dimensions.}} \\ \hline\hline
&  & TCA &  &  \\
$\alpha $ & $QSR_{\alpha }(+)$ & $QSR_{\alpha }(-)$ & $QSR_{\alpha }$ & $%
\delta _{\alpha }$ \\ \hline
1 & (\textbf{86.58}, 71.16) & (\textbf{-96.04}, -65.21) & \textbf{77.89} &
0.2524 \\
2 & (56.01, \textbf{61.84}) & (\textbf{-64.99}, -46.48) & \textbf{56.40} &
0.1041 \\
3 & (83.11, 41.49) & (-68.57, -54.06) & 57.79 & 0.0848 \\
4 & (65.59, 64.28) & (-69.82, -54.48) & 63.01 & 0.0701 \\ \hline\hline
&  & TLRA &  &  \\
$\alpha $ & $QSR_{\alpha }(+)$ & $QSR_{\alpha }(-)$ & $QSR_{\alpha }$ & $%
\delta _{\alpha }$ \\
1 & (\textbf{87.43},63.19) & (\textbf{-89.99},-50.36) & \textbf{68.69} &
61.9773 \\
2 & (47.57,\textbf{68.06}) & (\textbf{-62.51},-47.07) & \textbf{54.83} &
34.6618\\
3 & (71.49,61.51) & (-62.47,-62.94) & 64.37 & 32.4594 \\
4 & (60.55,51.41) & (-61.03,-52.59) & 56.05 & 20.4148\\ \hline
\end{tabular}

\begin{table}
\begin{tabular}{c}
  \includegraphics[width=19.3cm]{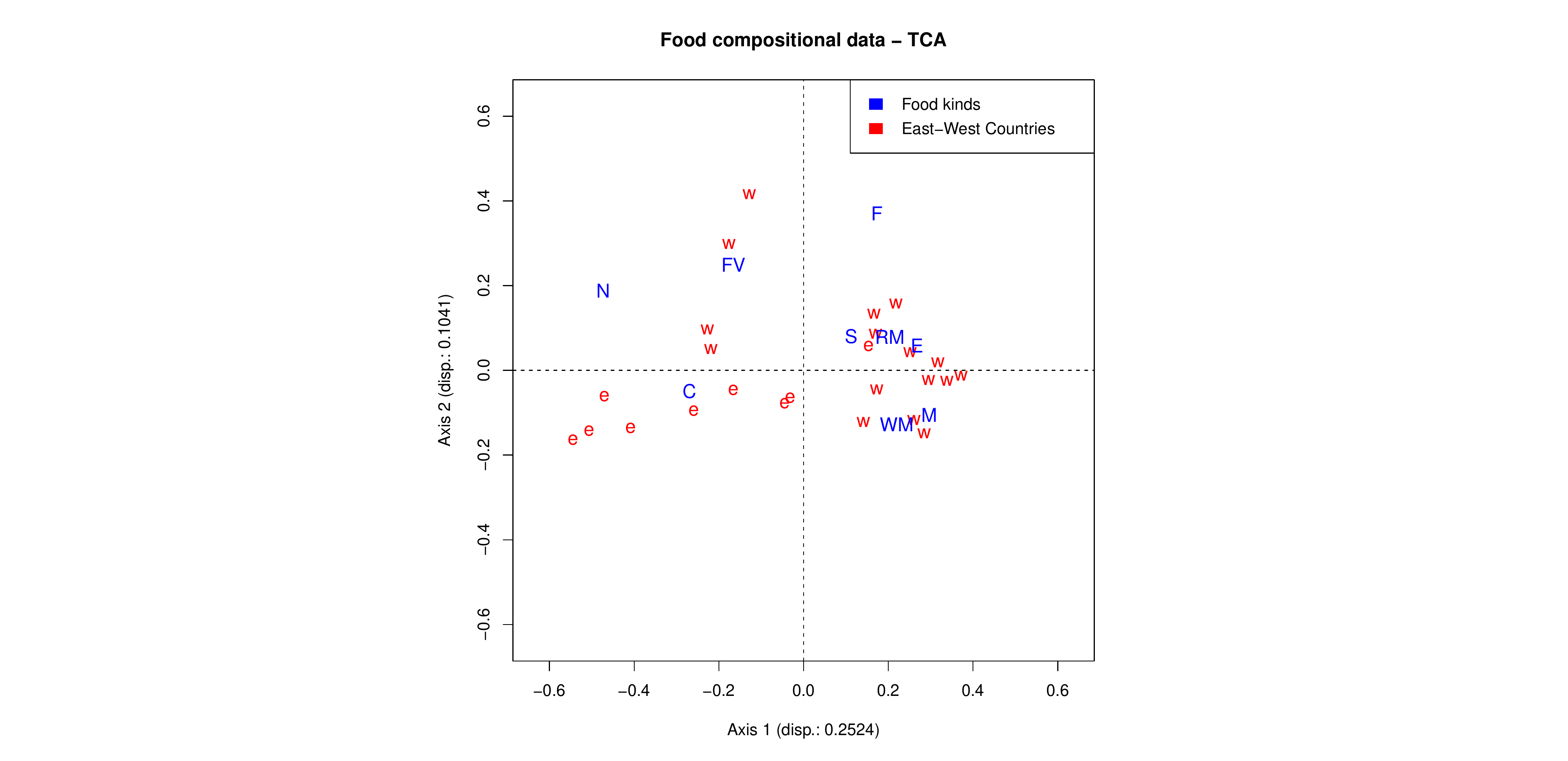}\\
 \includegraphics[width=19.3cm]{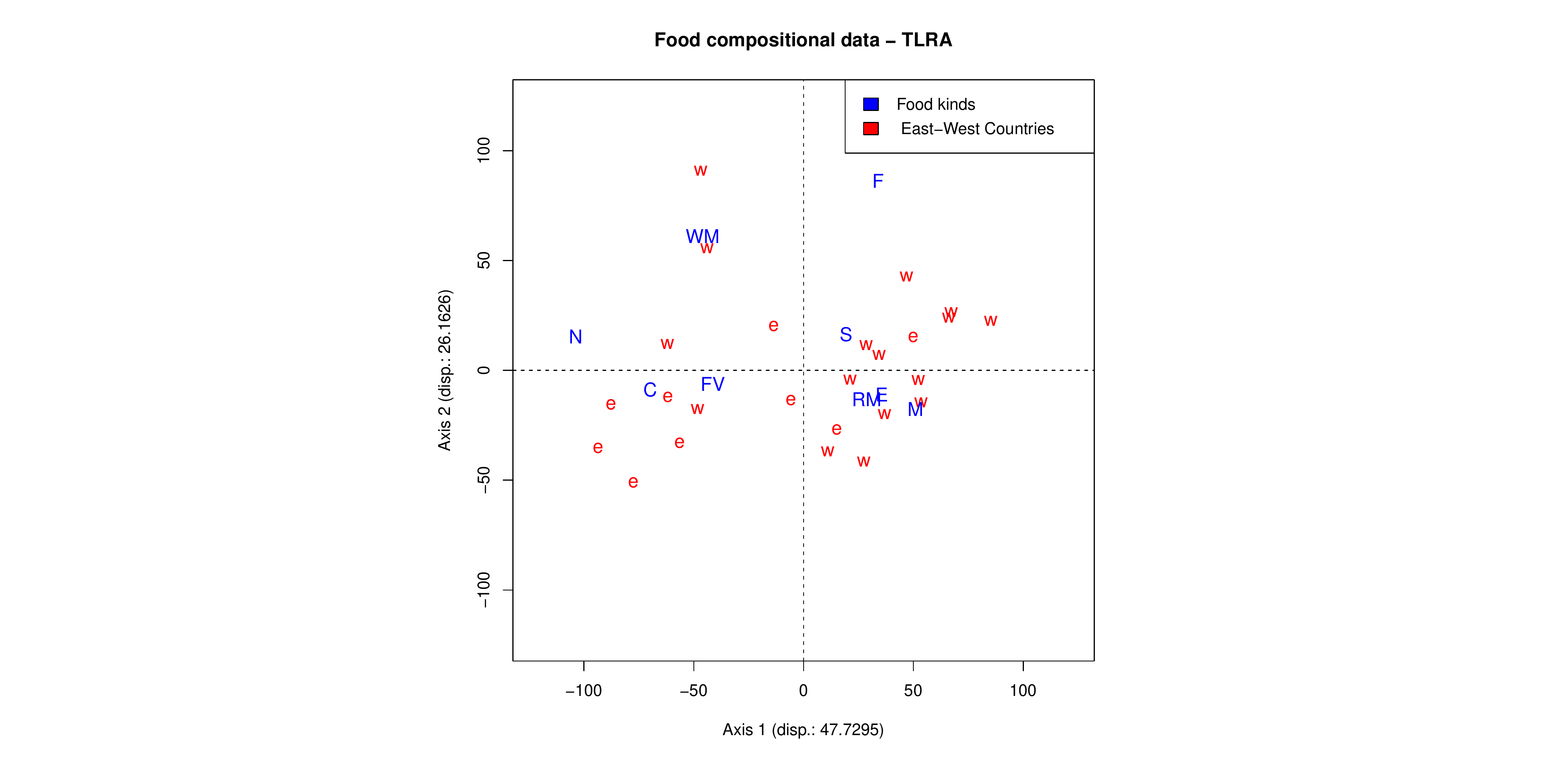}
\end{tabular}
\caption{\label{smith:fig3}  \textbf{ {\it Maps of Food compositional  data.}}}
\end{table}

\section{Minimal R script}

We show a minimal R script that can be executed to obtain the results in
this paper. A more extensive script will be available in the CRAN repository.

We follow the three steps procedure outlined in section 3. We suppose that
the data is in a matrix form.
\begin{itemize}
\item \textcolor{blue}{Step 1 (centering):  different for each method.}

\item \textcolor{blue}{Step 2 (computation of contribution scores and QSR index) : common to both
TCA and TLRA methods. This step calls the function TSVD.r from the R Package
TaxicabCA.}

\item \textcolor{blue}{Step 3 (visualisation). It  uses the plot function from TaxicabCA. For TCA, it
suffices to use the TaxicabCA plot function. For TRLA, we must create a
partial tca object in order to use the TaxicabCA plot function.}
\bigskip

\textcolor{green}{\#\#\# \textbf{Step 0: Preliminaries}}

\textcolor{green}{\# Install the 2 packages and the center\_scale function}
\begin{itemize}
\item{a.} \textcolor{green}{library(TaxicabCA)}

\item{b.}\textcolor{green}{library(GA)}

\item {c.} \textcolor{green}{center\_scale \TEXTsymbol{<}- function(x) \{
\textcolor{green}{scale(x, scale = FALSE)}}
\textcolor{green}{\}}
\end{itemize}
\bigskip
\end{itemize}
\# \textcolor{red}{dataMatrix holds the raw data}

\# 7x4 dataMatrix of Example 1

dataMatrix \textless \textendash  matrix(c(69,49,48,41,

148,45,14,22,

170,65,12,29,

159,57,12,28,

122,26,6,18,

106,21,5,23,

40,7,1,14),nrow=7,ncol=4,byrow=T)

rownames(dataMatrix) \textless \textendash  c(``16-24",``25-34",

``35-44",``45-54", ``55-64", ``65-74", ``75+")

colnames(dataMatrix) \textless \textendash  c(``Bad", ``Average", ``Good", ``VeryGood")

\# rownames and colnames are used to label points

dataName \textless \textendash ``XLStatCAData" \# Will appear in the figure
title\bigskip

\# Uncomment ONE of the following lines

\# to choose the centering method

\# centeringMethod \textless \textendash ``TCA"

centeringMethod \textless \textendash``TLRA"


\# ncol(centeredDataMatrix))

\# For this illustration, must have nAxes $\geq$ 2

nAxes \textless \textendash 2

dataMatrix  \textless \textendash as.matrix(dataMatrix)\bigskip

\#\#\#\#\#\#\#\#\#\#\#\#\#\#\#\#\#\#\#\#\#\#\#\#\#\#\#\#\#\#\#\#\#\#\#\#\#\#%
\#\#\#\#\#\#\#\#\#\#

\textbf{\#\#\# Step 1: Centering the data matrix:}

\# According to the centering method chosen ABOVE\bigskip

if (centeringMethod == ``TCA") \{

Proba \textless \textendash \, dataMatrix/sum(dataMatrix)

rowProba \textless \textendash \, apply(Proba,1,sum)

colProba \textless \textendash \, apply(Proba,2,sum)

centeredDataMatrix \textless \textendash \, Proba - rowProba \%*\% t(colProba)

\}

\# TLRA Centering

if (centeringMethod == ``TLRA") \{

centeredDataMatrix \textless \textendash \, log(dataMatrix)

centeredDataMatrix \textless \textendash \,scale(centeredDataMatrix, scale = FALSE)

centeredDataMatrix  \textless \textendash \, t(scale(t(centeredDataMatrix),scale =
FALSE))

attr(centeredDataMatrix,``scaled:center") \textless \textendash \, NULL

\}\bigskip

library(TaxicabCA)\bigskip

\textbf{\#\#\# Step 2: Compute the Taxicab SVD for the centered matrix}

\# Common to both methods

nRow \textless \textendash \, nrow(centeredDataMatrix)

nCol \textless \textendash \, ncol(centeredDataMatrix)

axesNames \textless \textendash \,paste(``Axis",1 :nAxes,sep=``")

\# Create the matrices required to receive the results

rowScores \textless \textendash \, matrix(NA, nrow = nRow, ncol = nAxes)

rownames(rowScores) \textless \textendash \, rownames(centeredDataMatrix)

colScores \textless \textendash \, matrix(NA, ncol = nCol, nrow = nAxes)

colnames(colScores) \textless \textendash \, colnames(centeredDataMatrix)

dispersion \textless \textendash \, rep(NA, nAxes) \# matrix(0, nrow = nAxes, ncol =
1)

QSR \textless \textendash \, matrix(NA, nrow = nAxes, ncol = 5)

colnames(QSR) \textless \textendash \, c(``VUQuadrant1",``VUQuadrant3", ``VUQuadrant2",``VUQuadrant4",``All")

rownames(QSR) \textless \textendash \,colnames(rowScores) \textless \textendash \,
rownames(colScores) \textless \textendash \,

names(dispersion) \textless \textendash \, axesNames

residuals  \textless \textendash \, centeredDataMatrix

iiAxis \textless \textendash \,1

for (iiAxis in 1 :nAxes) \{

\# The search functions come from TaxicabCA

\# Uncomment ONE search method - As of 2020, on a desktop computer,

\# Exhaustive is only feasible for nRow $<$  22

axisResult \textless \textendash \, SearchExhaustive(residuals)

\# axisResult \textless \textendash \,  SearchCrissCross(residuals)

\# axisResult \textless \textendash \,SearchGenteticAlgoritm(residuals)

\# Note: Some versions of TaxicaCA misspell ``Algoritm"!

U \textless \textendash \,axisResult\$uMax

dispersion[iiAxis] \textless \textendash \, axisResult\$L1Max

rowScores[, iiAxis]  \textless \textendash \, residuals \%*\% t(axisResult\$uMax)

V \textless \textendash \,sign(rowScores[, iiAxis, drop = F])

colScores[iiAxis, ] \textless \textendash \, t(V) \%*\% residuals

\# Compute the quality of the signs of

\# the residuals (QSR) for each UV ``quadrant"

QSR[iiAxis,1] \textless \textendash \, 0.25/sum(abs(residuals[V $>$ =0,U%
$>$=0]))

QSR[iiAxis,2] \textless \textendash \, 0.25/sum(abs(residuals[V $<$ 0,U
$<$ 0]))

QSR[iiAxis,4] \textless \textendash \, -0.25/sum(abs(residuals[V $>$= 0, U $ < $  0]))

QSR[iiAxis,3] \textless \textendash \, -0.25/sum(abs(residuals[V $<$0,U
$>$=0]))

\# Compute the overall quality of the signs of the residuals

\# abs(QSR[iiAxis,1:4])/(1/sum(abs(residuals)))

QSR[iiAxis,5]  \textless \textendash \,  1/sum(abs(residuals))

QSR[iiAxis,]  \textless \textendash \, dispersion[iiAxis]*QSR[iiAxis,]

\# Update the residuals for the next iteration

residuals  \textless \textendash \, residuals - (rowScores[, iiAxis, drop = F] \%*\%

colScores[iiAxis,,drop = F])/dispersion[iiAxis]

\}

\bigskip

\textbf{\#\#\# Step 3: Visualisation}

\# TCA Visualisation

if (centeringMethod == ``TCA") \{

\# tca can choose a search method automatically

Data.tca  \textless \textendash \, tca(dataMatrix, nAxes=nAxes)

Data.tca\$dataName  \textless \textendash \,  paste(dataName,``TCA",sep= `` - ")

\# Open a graphics window outside of RStudio (if RStudio is used)

dev.new(noRStudioGD = TRUE)

\# Call plot.tca from TaxicabCA (Data.tca is class ``tca")

plot(Data.tca,labels.rc = c(1, 1),cex.rc = c(.8,.8))

\}

\# TLRA Visualisation

if (centeringMethod == ``TLRA") \{

rowScores  \textless \textendash \,  nRow*rowScores

colScores  \textless \textendash \, nCol*colScores

Data.tlra  \textless \textendash \, list(rowScores = rowScores, colScores = colScores,

dispersion = dispersion,

dataName = paste(dataName,``TRLA",sep= `` - "))

\# Add class ``tca" to the class of the list in order to

\# call plot.tca from TaxicabCA automatically

class(Data.tlra)  \textless \textendash \,  c(class(Data.tlra),``tca")

\# Open a graphics window outside of RStudio (if RStudio is used)

dev.new(noRStudioGD = TRUE)

\# Call plot.tca from TaxicabCA - Data.tlra is class ``tca"

\# Use labels.rc = c(1,1) only if the data has rown and colnames.

\# Otherwise, use labels.rc = c(0,0), c(1,0) or c(0,1)

plot(Data.tlra,labels.rc = c(1,1),cex.rc = c(.8,.8))

\}\bigskip

\# Print the numerical results

print(dataName)

print(centeringMethod)

print(dispersion)

print(QSR)

\section{Conclusion}

We attained two aims in this paper. First, we introduced the package
TaxicabCA in R and showed its functionalities. Second, we applied it to
study the influence of two different centering procedures in two well
developed methods CA and LRA. In both aims, the tool was TSVD, which has
some nice mathematical properties.

In this paper we only considered unweighted LRA analysis as developed in
Goodman (1991, equation 2.2.1) or Aitchison (1986). We did not tackle
weighted LRA analysis as discussed in Goodman (1991, equation 2.2.7) and
Greenacre and Lewi (2009), because the weighted residual matrix is not
uniformly centered, so TSVD will produce biased results because equations
(15,16) wil not be satisfied.

There are two perspectives for data analysis of contingency tables or
compositional data. The first, based on the mathematical fact that the rows
or the columns are found on the probability simplex, is CA/LRA based on
invariance principles: such as principle of distributional equivalence, and,
the principle of scale invariance and the principle of subcompositional
coherence. The second is on re-expression and the analysis of residuals
advocated by Tukey (1977, in particular chapters 10 and 15). This paper
develops jointly the use of both perspectives by introducing the use of QSR
index describing the concentration of the sign of the residuals in
re-expressed count or compositional data, and thus choosing the method. We
think our novel approach will especially be fruitful for large data sets,
such as microbiome data.

We conclude by citing Tukey (1977, p.400): \textquotedblright \textit{the
general maxim--it is a rare thing that a specific body of data tells us
clearly enough how it itself should be analyzed--applies to choice of
re-expression for two-way analysis}\textquotedblright . For contingency
tables and compositional data, the choice of re-expression is essentially
between equations (11) and (14).

\bigskip
\begin{verbatim}
Acknowledgements.
\end{verbatim}

Choulakian's research has been supported by NSERC of Canada.\bigskip

\textbf{References}

\bigskip

Aitchison, J. (1986). \textit{The Statistical Analysis of Compositional Data}%
. London: Chapman and Hall.

Allard, J. and Choulakian, V. (2019). TaxicabCA in R, available on CRAN.

Beh,  E. J. and  Lombardo, R.  (2014).  \textit{Correspondence Analysis: Theory, Practice and New Strategies.} UK: John Wiley $\&$ Sons.

Benz\'{e}cri, J.P. (1973).\ \textit{L'Analyse des Donn\'{e}es: Vol. 2:
L'Analyse des Correspondances}. Paris: Dunod.

Choulakian, V. (2006). Taxicab correspondence analysis. \textit{%
Psychometrika,} 71, 333-345.

Choulakian, V. (2008). Taxicab correspondence analysis of contingency tables
with one heavyweight column. \textit{Psychometrika}, 73(2), 309-319.

Choulakian, V., Simonetti, B. and Gia, T.P. (2014). Some further aspects of
taxicab correspondence analysis. \textit{Statistical Methods and Applications%
}, 23, 401-416.

Choulakian, V. (2016). Matrix factorizations based on induced norms. \textit{%
Statistics, Optimization and Information Computing}, 4, 1-14.

Choulakian, V. (2017). Taxicab correspondence analysis of sparse contingency
tables. \textit{Italian Journal of Applied Statistics,} 29 (2-3), 153-179.

Choulakian, V. and Abou-Samra, G. (2020). Mean absolute deviations about the
mean, the cut norm and taxicab correspondence analysis. \textit{Open Journal
of Statistics}, 10(1), 97-112

Genet, J.P. (2002). Analyse factorielle et construction des variables.
L'origine g\'{e}ographique des auteurs anglais (1300-1600). \textit{Histoire
\& Mesure}, XVII (1/2), 1-19.

Goodman, L.A. (1991) Measures, models, and graphical displays in the
analysis of cross-classified data. \textit{Journal of the American
Statistical Association}, 86 (4), 1085-1111.

Goodman, L.A. (1996). A single general method for the analysis of
cross-classified data: Reconciliation and synthesis of some methods of
Pearson, Yule, and Fisher, and also some methods of correspondence analysis
and association analysis.\textit{\ Journal of the American Statistical
Association}, 91, 408-428.

Greenacre, M. and Lewi, P. (2009). Distributional equivalence and
subcompositional coherence in the analysis of compositional data,
contingency tables and ratio-scale measurements. \textit{Journal of
Classification}, 26, 29-54.

Tukey, J.W. (1977). \textit{Exploratory Data Analysis}. Addison-Wesley:
Reading, Massachusetts.

\end{document}